\documentclass[journal]{IEEEtran}
\ifCLASSINFOpdf
\else
\fi
\usepackage{graphicx}
\usepackage{cite}
\usepackage{amsmath}
\usepackage{amsfonts}
\usepackage{enumerate}
\usepackage{algpseudocode}
\usepackage{algorithmicx,algorithm}
\usepackage{bm}

\begin{document}
%
\title{Sphere Decoder with Box Optimization for \\FTN Non-orthogonal FDM System}
%
%
%

\author{Mengqi Guo$^*$, Ji Zhou$^*$, Yueming Lu, and Yaojun Qiao
\thanks{This work was supported in part by the National Natural Science Foundation of China under Grant 61771062 and Grant 61427813; National Key Research and Development Program under Grant 2016YFB0800302; BUPT Excellent Ph.D. Students Foundation CX2017207. (\emph{Corresponding author: Yaojun Qiao})}
\thanks{Mengqi Guo, Ji Zhou, and Yaojun Qiao are with the Beijing Key Laboratory of Space-ground Interconnection and Convergence, School of Information and Communication Engineering, Beijing University of Posts and Telecommunications (BUPT), Beijing 100876, China (e-mail: qiao@bupt.edu.cn).}
\thanks{Yueming Lu is with the Key Laboratory of Trustworthy Distributed Computing and Service, Ministry of Education, School of Cyberspace Security, Beijing University of Posts and Telecommunications (BUPT), Beijing 100876, China.}
\thanks{* The authors contribute the same for this manuscript.}
}

\maketitle

\begin{abstract}
In 1975, the pioneering work of J. E. Mazo showed the potential faster-than-Nyquist (FTN) gain of single-carrier binary signal. If the inter-symbol interference is eliminated by an optimal detector, the FTN single-carrier binary signal can transmit 24.7\% more bits than the Nyquist signal without any loss of bit error rate performance, which is known as the Mazo limit. In this paper, we apply sphere decoder (SD) with box optimization (BO) to reduce inter-carrier interference (ICI) in FTN non-orthogonal frequency division multiplexing (FTN-NOFDM) system. Compared with the conventional SD, SD with BO can achieve the same performance to reduce ICI, but its average number of expanded nodes in search process is significantly decreased especially for high-order modulation format, which can reduce the complexity of the receiver. When the bandwidth compression factor $\bm{\alpha}$ is set to 0.802, the transmission rate of QPSK-modulated FTN-NOFDM is 24.7\% faster than the Nyquist rate, and it has almost the same performance as orthogonal frequency division multiplexing (OFDM), which agrees well with the Mazo limit. The QPSK-modulated FTN-NOFDM with $\bm{\alpha}$ equal to 0.5 (the spectral efficiency is 4 bit/s/Hz) outperforms 16QAM-modulated OFDM by about 1.5 dB. The 16QAM-modulated FTN-NOFDM with $\bm{\alpha}$ equal to 0.67 and 0.5 (the spectral efficiency is 6 bit/s/Hz and 8 bit/s/Hz, respectively) outperforms 64QAM-modulated and 256QAM-modulated OFDM by about 1.5 dB and 2 dB, respectively. Therefore, FTN-NOFDM will be a promising modulation scheme for the future bandwidth-limited wireless communications.
\end{abstract}


%
\IEEEpeerreviewmaketitle

\section{Introduction}
\IEEEPARstart{W}{ith} the increasing demand on transmission capacity, faster-than-Nyquist (FTN) signaling has attracted lots of attention from the research community due to its improvement in spectral efficiency for single-carrier and multi-carrier signal \cite{FTN1,FTN2,FTN3,FTN4}. It has become a potential candidate for the fifth-generation (5G) communications \cite{5G1,5G2,5G3}. In 1975, J. E. Mazo first proposed the potential FTN gain of single-carrier binary signal \cite{Mazo}. It is revealed that when the time acceleration factor is set between 1 and 0.802, the inter-symbol interference can be eliminated by an optimal detector. Under this circumstance, the FTN single-carrier binary signal can transmit up to 24.7\% more bits than the Nyquist signal without any loss of bit error rate (BER) performance, which is known as the Mazo limit. In recent decades, the related research on single-carrier FTN signaling continues to spring up \cite{SC1,SC2,SC3,SC4,SC5}.

To improve the spectral efficiency of multi-carrier signal, spectrally efficient frequency division multiplexing (SEFDM) is proposed \cite{SEFDM1,SEFDM2,SEFDM3}. As a fractional Fourier transform-based (FrFT-based) non-orthogonal frequency division multiplexing (NOFDM) scheme, the spectral efficiency of SEFDM is increased by compressing the subcarrier spacing. The subcarrier spacing of discrete Fourier transform-based (DFT-based) orthogonal frequency division multiplexing (OFDM) is the symbol rate per subcarrier, while that of SEFDM is less than the symbol rate per subcarrier. The orthogonality between subcarriers is damaged due to compression of subcarrier spacing, which induces inter-carrier interference (ICI) between subcarriers, and the ICI becomes larger with the increasing of compression. Therefore, the methods to generate subcarrier compressed multi-carrier signal with less ICI and effectively reduce ICI at receiver are key points to pay attention to.

Recently, the discrete cosine transform-based (DCT-based) OFDM has received extensive attention. The subcarrier spacing of DCT-based OFDM is half of the symbol rate per subcarrier, and the subcarriers are still orthogonal to each other \cite{DCT1,DCT2,DCT3}. For DFT-based OFDM, the orthogonality between subcarriers is damaged when the subcarrier spacing is less than the symbol rate per subcarrier, but for DCT-based OFDM it is damaged when the subcarrier spacing is less than half of the symbol rate per subcarrier. The FTN-NOFDM signal is generated when the subcarrier spacing is less than half of the symbol rate per subcarrier \cite{FrHT-FTN,FrCT-FTN}. Therefore, under the same spectral efficiency the fractional cosine transform-based (FrCT-based) FTN-NOFDM signal has less ICI than FrFT-based FTN-NOFDM signal \cite{FrCT-FTN}. The investigation of this paper is based on FrCT-based FTN-NOFDM signal.



\begin{figure*}[!t]
\centering
\includegraphics[width=16.5cm]{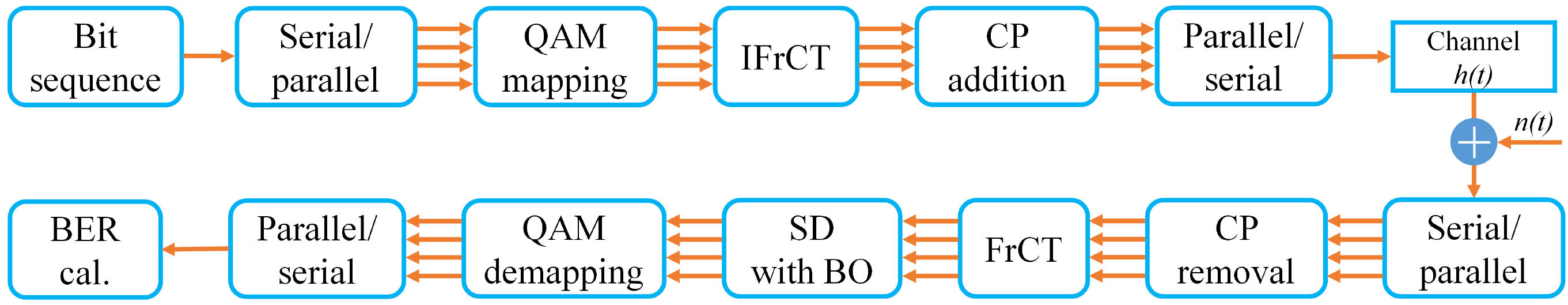}
\caption{Block diagram of DCT-based FTN-NOFDM system. QAM: quadrature amplitude modulation; IFrCT: inverse fraction cosine transform; CP: cyclic prefix; SD: sphere decoder; BO: box optimization; BER Cal.: bit error rate calculation.}
\label{fig1}
\end{figure*}

In FrCT-based FTN-NOFDM system, ICI seriously degrades the system performance due to the non-orthogonality between subcarriers. The NOFDM system with $N$ subcarriers can be considered as an $N \times N$ multiple-input multiple-output (MIMO) system \cite{MIMO}, so that some MIMO detectors to reduce interference between antennas also have potential to reduce ICI between subcarriers. In MIMO system, the maximum likelihood detector (MLD) is recognized as an optimal detector, but it is impossible to be used in practical system owing to the high complexity. Sphere decoder (SD) can achieve maximum likelihood performance with lower complexity than MLD \cite{SD1,SD2}, but when the high-order modulation format is applied, the expected complexity is exponential as well \cite{SD_complexity}. Therefore, the complexity of SD still has possibility to be further reduced.

There are a number of methods to reduce the complexity of SD, such as semi-definite relaxation \cite{SDR}, geometrical method \cite{geometrical}, increasing radius SD \cite{increasing radius}, K-best SD\cite{I-K-best}, statistical pruning SD \cite{statistical_pruning}, SD-based sequence estimation \cite{SDSE} and so on. In this paper, we apply a low-complexity SD with box optimization (BO) \cite{BO} to reduce ICI in FTN-NOFDM system. The main contributions of this paper are listed as follows:

\begin{itemize}
\item SD with BO can reduce the average number of expanded nodes of the conventional SD in the search process, and it still has entirely the same performance to reduce ICI as the conventional SD in FTN-NOFDM. The low-complexity property of SD with BO is more obvious in FTN-NOFDM with high-order constellation.
\item When the bandwidth compression factor $\alpha$ is set to 0.802, the simulation results show that the transmission rate of QPSK-modulated FTN-NOFDM is 24.7\% faster than the Nyquist rate, and it has almost the same performance as OFDM, which agrees well with the Mazo limit.
\item The QPSK-modulated or 16QAM-modulated FTN-NOFDM has better BER performance than OFDM with high-order constellation under the same spectral efficiency. When the spectral efficiency is 4 bit/s/Hz, QPSK-modulated FTN-NOFDM outperforms 16QAM-modulated OFDM by about 1.5 dB. When the spectral efficiency is 6 bit/s/Hz and 8 bit/s/Hz, respectively, 16QAM-modulated FTN-NOFDM outperforms 64QAM-modulated and 256QAM-modulated OFDM by about 1.5 dB and 2 dB, respectively.
\end{itemize}

The remainder of this paper is organized as follows. Section \ref{section2} introduces the principle of generating FrCT-based FTN-NOFDM signal. The details about the search process of the conventional SD are described in Section \ref{section3}. In Section \ref{section4}, the search process of SD with BO and the details of BO algorithm are described. Section \ref{section5} presents the simulation results. Finally, the paper is concluded in Section \ref{section6}.


\section{Principle of FTN-NOFDM System} \label{section2}

\begin{figure}[!t]
\centering
\includegraphics[width=8.6cm]{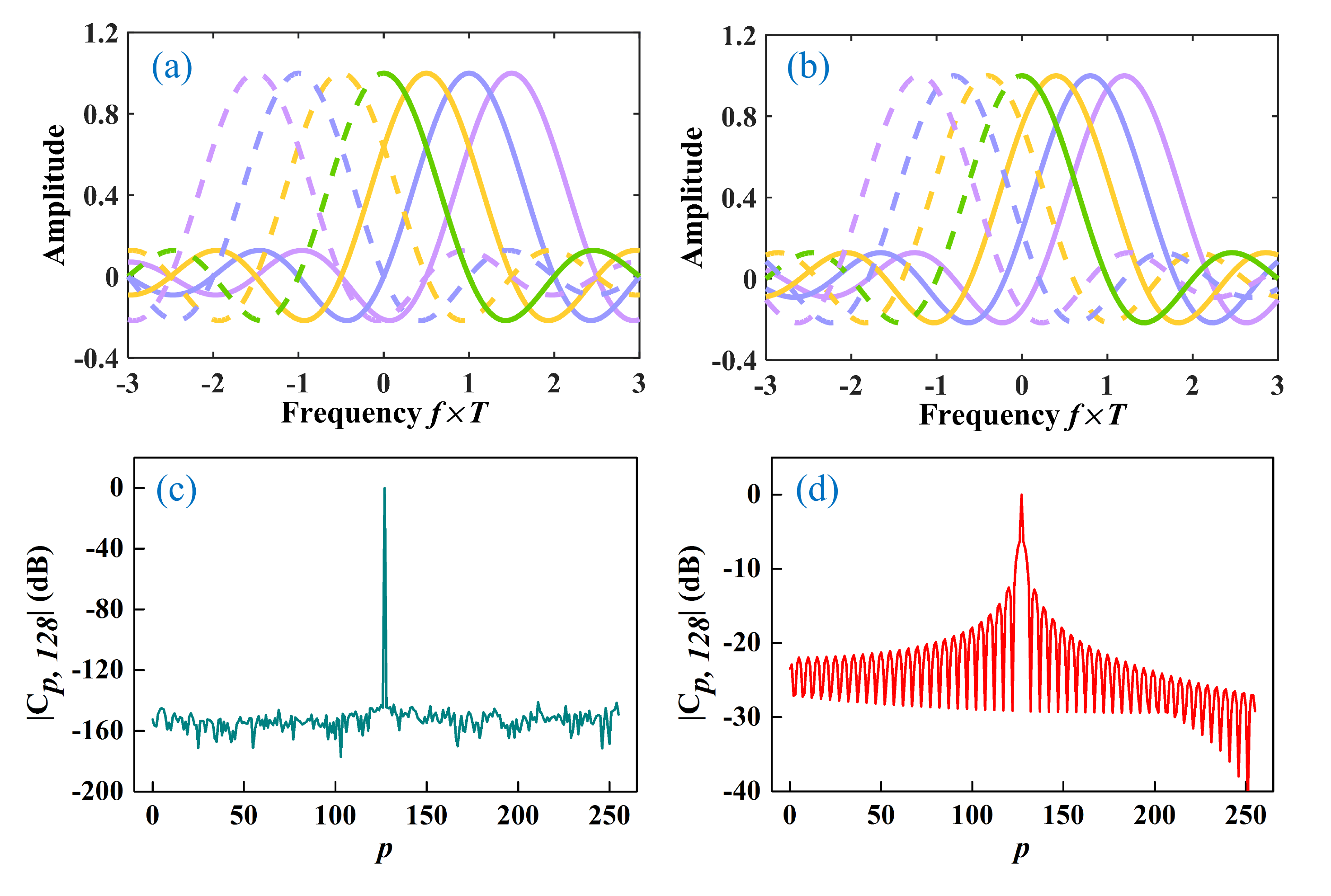}
\caption{(a) Sketched spectrum of DCT-based OFDM; (b) Sketched spectrum of FrCT-based FTN-NOFDM system with $\alpha=0.8$; (c) $C_{p,128}$ against $p$ for DCT-based OFDM; (d) $C_{p,128}$ against $p$ for FrCT-based FTN-NOFDM system with $\alpha=0.8$.}
\label{fig2}
\end{figure}

The block diagram of FTN-NOFDM system based on FrCT is demonstrated in Fig.~\ref{fig1}. The $N$-point inverse FrCT (IFrCT) and FrCT are defined as
\begin{equation}
{x_n} = \sqrt {\frac{2}{N}} \sum\limits_{k = 0}^{N - 1} {{W_k}{X_k}\cos \left( {\frac{{\alpha\pi (2n + 1)k}}{{2N}}} \right),}
\label{eq1}
\end{equation}
\begin{equation}
{X_k} = \sqrt {\frac{2}{N}} {W_k}\sum\limits_{n = 0}^{N - 1} {{x_n}\cos \left( {\frac{{\alpha\pi (2n + 1)k}}{{2N}}} \right),}
\label{eq2}
\end{equation}
where $0 \le n \le N - 1$, $0 \le k \le N - 1$, and
\begin{equation}
{W_k} = \left\{ \begin{array}{l}
\frac{1}{{\sqrt 2 }},{\kern 5pt} k = 0\\
1,{\kern 13pt} k = 1,2, \ldots ,N - 1.
\end{array} \right.
\label{eq3}
\end{equation}

In Eqs. (\ref{eq1}) and (\ref{eq2}), $\alpha$ means the bandwidth compression factor. As shown in Fig. \ref{fig2} (a), the OFDM signal based on DCT is generated when $\alpha$ is equal to $1$, and the subcarrier spacing is half of the symbol rate per subcarrier. All the subcarriers fall on the positive frequencies, and the subcarriers on negative frequencies are just the corresponding images of positive frequencies. Fig. \ref{fig2} (b) depicts the sketched spectrum of FrCT-based FTN-NOFDM with $\alpha$ of 0.8 as an example. The subcarrier spacing of FrCT-based FTN-NOFDM is compressed to less than half of the symbol rate per subcarrier. The baseband bandwidth is smaller than that of DCT-based OFDM with 20\% bandwidth saving, and the transmission rate is 25\% faster than the Nyquist rate. Therefore, the FrCT-based FTN-NOFDM signal is generated when $\alpha$ is less than 1.

The compression of subcarrier spacing leads to the non-orthogonality between subcarriers, which introduces ICI into the FTN-NOFDM system. The ICI between subcarrier $p$ and subcarrier $q$ is represented by cross-correlation value ${C_{p,q}}$ in the $N\times N$ correlation matrix ${\bm{C}}$ as
\begin{equation}
\begin{aligned}
{C_{p,q}} = \frac{2}{N}\sum\limits_{n = 0}^{N - 1} &{W_p\text{cos}\left( {\frac{{\alpha \pi (2n+1)p}}{2N}} \right)}\\
&\times W_q\text{cos}\left( {\frac{{\alpha \pi (2n+1)q}}{2N}} \right).
\end{aligned}
\label{eq4}
\end{equation}
Fig. \ref{fig2} (c) depicts the $C_{p,128}$ against $p$ for DCT-based OFDM. The correlation matrix C is an identity matrix for $\alpha=1$, which means the subcarriers of DCT-based OFDM are orthogonal. Fig. \ref{fig2} (d) shows the $C_{p,128}$ against $p$ for FrCT-based FTN-NOFDM with $\alpha$ of 0.8. The cross-correlation values in $\bm{C}$ are not zero once $\alpha$ is less than 1, which means the subcarriers of FrCT-based FTN-NOFDM are no longer orthogonal to each other. The introduced ICI between subcarriers can seriously degrade the system performance.

\section{Sphere Decoder} \label{section3}
\begin{figure}[!t]
\centering
\includegraphics[width=7cm]{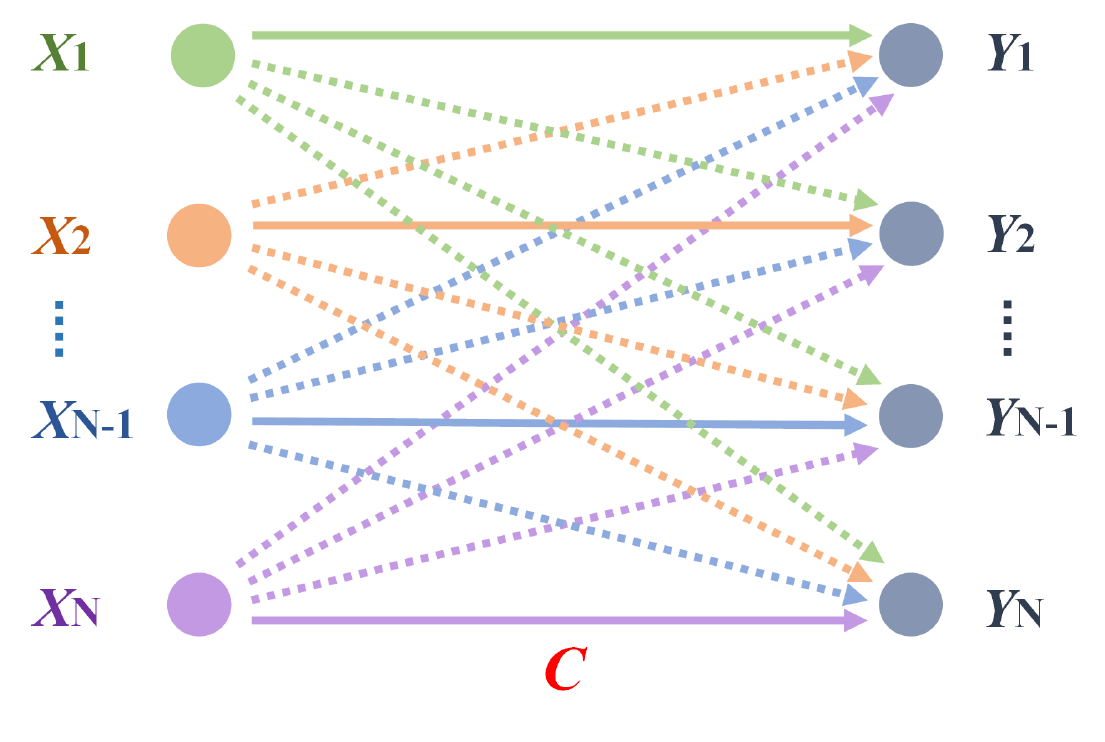}
\caption{MIMO system model for FTN-NOFDM. The correlation matrix $\bm{C}$ of FTN-NOFDM can be considered as channel matrix $\bm{H}$ of MIMO.}
\label{fig3}
\end{figure}

Considering the whole modulation and demodulation of FTN-NOFDM system, the $N$-dimensional signal ${\bm{X}}$ inputs into the $N$-point IFrCT at transmitter, and $N$-dimensional signal ${\bm{Y}}$ outputs from $N$-point FrCT at receiver. The recovered signal ${\bm{Y}}$ can be defined as
\begin{equation}
{\bm{Y}} = {\bm{CX}} + {\bm{G}},
\label{eq5}
\end{equation}
where ${\bm{G}}$ is additive white Gaussian noise (AWGN). As shown in Fig. \ref{fig3}, the FTN-NOFDM with $N$ subcarriers can be regarded as an $N \times N$ MIMO system. The $N$-dimensional input signal $\bm{X}$ and the $N$-dimensional output signal $\bm{Y}$ can be considered as $N$ transmitter antennas and $N$ receiver antennas, respectively. The correlation matrix ${\bm{C}}$ with interference between subcarriers is equivalent to the channel matrix $\bm{H}$ with interference between antennas, which can be accurately calculated from Eq. (\ref{eq4}). Consequently, some effective detectors to reduce interference between antennas in MIMO system also have potential to reduce ICI in FTN-NOFDM system.

The widely known optimal detector is the MLD, in which the symbols on $N$ subcarriers are determined by solving the following least-square problem \cite{MLD},
\begin{equation}
{\bm{X}_{MLD}} = \mathop {\arg \min }\limits_{{\bm{X}}\in {\mathcal{D}^N}} {\left\| {{\bm{Y}} - {\bm{CX}}} \right\|^2} = \mathop {\arg \min }\limits_{{\bm{X}}\in {\mathcal{D}^N}} E({\bm{X}}),
\label{eq6}
\end{equation}
where $\mathcal{D}$ is the constellation set, $E(\bm{X})$ means the metric value of the whole path. Eq. (\ref{eq6}) reveals that the exponential complexity of MLD makes it unrealizable for practical FTN-NOFDM system when the high-order modulation format is used. Compared with MLD, SD gives the maximum likelihood performance with reduced complexity. SD only searches constellation points lying within a $N$-dimensional hyperspace sphere of radius $\beta$, so the solution of SD should satisfy
\begin{equation}
{\bm{X}_{SD}} = \mathop {\arg \min }\limits_{{\bm{X}}\in {\mathcal{D}^N}} \left\{ {E({\bm{X}})\le \beta} \right\}.
\label{eq7}
\end{equation}

The QR decomposition is often applied to the correlation matrix $\bm{C}$ as $\bm{C}=\bm{Q}\bm{R}$, by this way the complexity of calculating the metric can be reduced. The $N \times N$ matrix $\bm{Q}$ is an unitary matrix, and the $N \times N$ matrix $\bm{R}$ is an upper triangular matrix. The matrix $\bm{R}$ is represented as
\begin{equation}
{\bm{R}} = \left[ {\begin{array}{*{20}{c}}
{{R_{0,0}}}&{{R_{0,1}}}& \cdots &{{R_{0,N - 1}}}\\
0&{{R_{1,1}}}& \cdots &{{R_{1,N - 1}}}\\
 \vdots &{}& \ddots & \vdots \\
0& \cdots &0&{{R_{N - 1,N - 1}}}
\end{array}} \right].
\label{eq8}
\end{equation}
With QR decomposition, the path metric in the search process can be calculated as the accumulation of the metric of each branch path. Without QR decomposition, each time the next node is searched, the path metric is no longer the accumulation metric of each branch path, and it needs to be recalculated for the whole path. Therefore, the QR decomposition is applied to get ${\bm{Z}} = {\bm{Q}^{T}}{\bm{Y}}$, and $E(\bm{X})$ can be calculated as
\begin{equation}
{E(\bm{X})} = {\left\| {{\bm{Z} - \bm{RX}}} \right\|^2}\le \beta.
\label{eq9}
\end{equation}


Then, the search process starts from subcarrier $N-1$ and moves to subcarrier $0$ owing to the upper triangular structure of matrix $\bm{R}$. The metric ${E(\bm{X})}$ can be expressed as the accumulation of each branch metric in the search process,
\begin{equation}
\begin{aligned}
{E(\bm{X})} &= {\left| {Z_0} - \sum\limits_{j = 0}^{N - 1} {{R_{0,j}}{X_j}} \right|^2}+{\left| {Z_1} - \sum\limits_{j = 1}^{N - 1} {{R_{1,j}}{X_j}} \right|^2}+\dots\\
&+{\left| {Z_{N-1}} - {R_{N-1,N-1}}{X_{N-1}} \right|^2}\\
&=E(X_0)+E(X_1)+\dots+E(X_{N-1}).
\end{aligned}
\label{eq10}
\end{equation}
After making decision for subcarrier $k$, the vector ${{\bm{X}}_{k:N - 1}}$ ($k:N-1$ means $k,k+1,\dots,N-1$) has been assigned decision, and the vector ${{\bm{X}}_{0:k - 1}}$ has not been assigned. The metric ${E(\bm{X})}$ can be expressed as the summation of ${E(\bm{X}_{0:k-1})}$ and ${E(\bm{X}_{k:N-1})}$ that
\begin{equation}
\begin{array}{l}
{E(\bm{X})}=E(\bm{X}_{0:k-1})+E(\bm{X}_{k:N-1})\\
=\left\| {{{\bf{Z}}_{0:k - 1}} \!-\! {{\bf{R}}_{0:k - 1,0:k - 1}}{{\bf{X}}_{0:k - 1}} \!-\! } \right.{{\bf{R}}_{0:k - 1,k:N - 1}}{\left. {{{\bf{X}}_{k:N - 1}}} \right\|^2}\\
+{\left\| {{{\bm{Z}}_{k:N - 1}} \!-\! {{\bm{R}}_{k:N - 1,k:N - 1}}{{\bm{X}}_{k:N - 1}}} \right\|^2}.
\end{array}
\label{eq11}
\end{equation}
The first term $E({\bf{X}}_{0:k - 1})$ of Eq. (\ref{eq11}) is commonly neglected in SD, and the bound $\beta$ is applied to $E({\bf{X}}_{k:N - 1})$,
\begin{equation}
{E({\bm{X}}_{k:N - 1})} \!=\! {\left\| {{{\bm{Z}}_{k:N - 1}} \!-\! {{\bm{R}}_{k:N - 1,k:N - 1}}{{\bm{X}}_{k:N - 1}}} \right\|^2} \le \beta.
\label{eq12}
\end{equation}
Therefore, the node of subcarrier $k$ is expanded only if the Eq. (\ref{eq12}) is satisfied. If Eq. (\ref{eq12}) is not satisfied, the current node and all of its child nodes are discarded. The search process for SD is illustrated as follows \cite{process},

\begin{enumerate}[1)]
\item Preprocessing: Start from the subcarrier $k=N-1$ and set the initial bound $\beta_{best}=\infty$.
\item Finite-State Machine (FSM):
\begin{enumerate}[a)]
\item If there exists any node that has not been searched at subcarrier $k$, choose one node and calculate branch metric $E(X_k)$, set $\beta_{new}=E({\bm{X}}_{k:N - 1})$.
\begin{itemize}
\item If $\beta_{new}< \beta_{best}$ and $k>0$, go to State A;
\item If $\beta_{new}< \beta_{best}$ and $k=0$, go to State B;
\item If $\beta_{new} \ge \beta_{best}$, go to State C.
\end{itemize}
\item If all the nodes have been searched at subcarrier $k$.
\begin{itemize}
\item If $k=N-1$, exit it;
\item If $k\ne N-1$, move to subcarrier $k+1$ and go back to FSM.
\end{itemize}
\end{enumerate}

\item State A: Move to subcarrier $k-1$ and go back to FSM.
\item State B: Record $\beta_{best}=\beta_{new}$ and store the decision $\bm{X}$ for all the subcarriers. If all the nodes have been searched for $k=0$, move to $k=1$ and go back to FSM; else keep in $k=0$ and go back to FSM.
\item State C: If all the nodes have been searched for subcarrier $k$, move to subcarrier $k+1$ and go back to FSM; else keep in subcarrier $k$ and go back to FSM.
\end{enumerate}

In the search process, the enumeration method should be properly chosen to improve the search efficiency. In other words, each node represents a constellation point, which node is first chosen influences the complexity of SD. The Fincke-Pohst (F-P) enumeration and Schnorr-Euchner (S-E) enumeration are two widely known enumeration methods \cite{Closest,Su}. For F-P enumeration, the nodes are searched just from left to right in nature order\cite{FP}. For S-E enumeration, the nodes are searched in the increasing order of their metrics \cite{SE}. Once a decision $\bm{X}$ is obtained, the radius $\beta$ is reduced, so only the nodes with smaller metric are expanded in the next step. The S-E enumeration discovers the eligible decision $\bm{X}$ more quickly than F-P enumeration, and the number of expanded nodes for S-E enumeration is less than that of F-P enumeration. Therefore, the S-E enumeration is used for the search process of SD, and no initial bound $\beta$ is required \cite{Closest}.

\begin{figure}[!t]
\centering
\includegraphics[width=8cm]{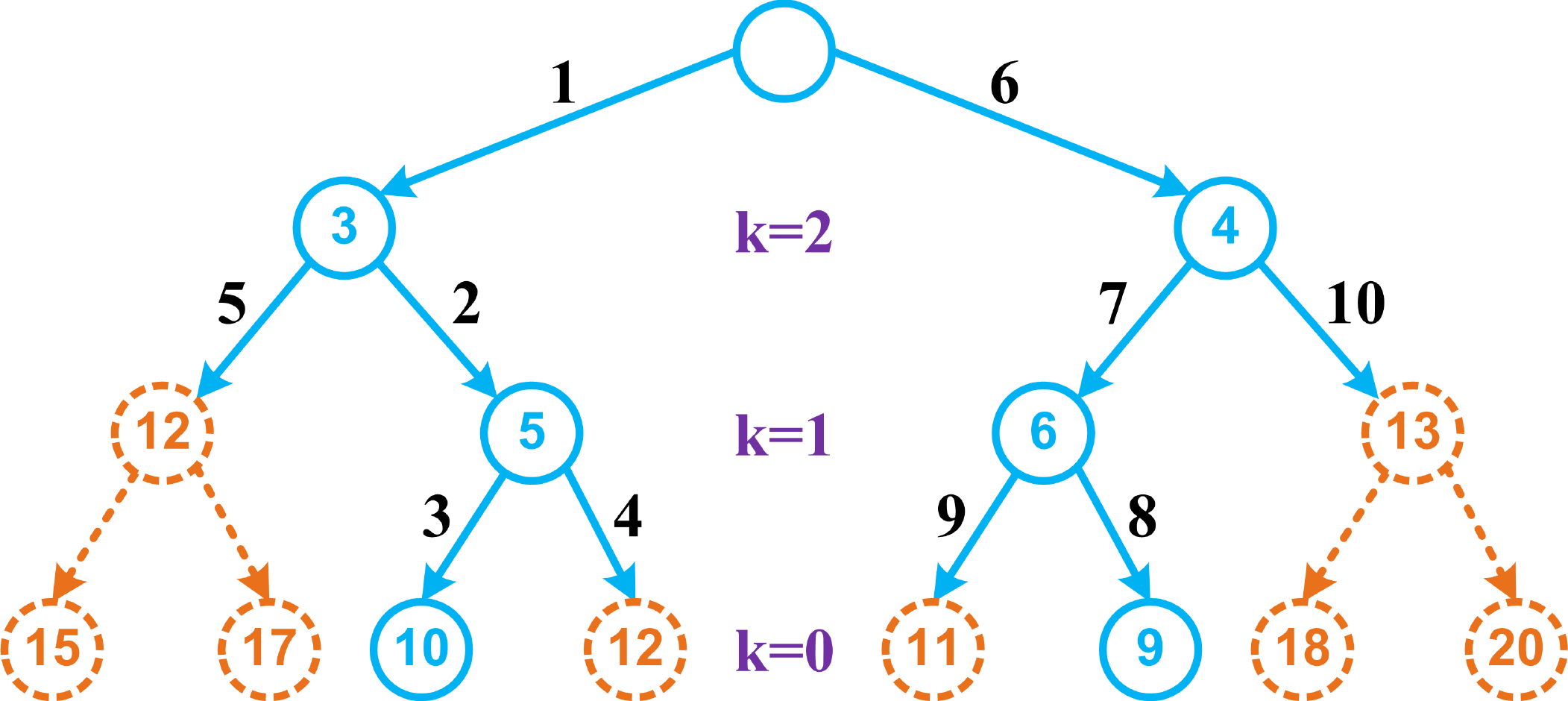}
\caption{An example of the search process for SD with BPSK modulation and $N=3$. The number on the branch path means the search order. The number on the node means the accumulated partial metric.}
\label{fig4}
\end{figure}
An example of SD with BPSK modulation and $N=3$ is illustrated in Fig. \ref{fig4} to show the search process more clearly. The number on the branch path means the search order. The number on the node means the accumulated partial metric. In the third step the first decision $\bm{X}$ is obtained, which is commonly called the Babai point \cite{Babai}. The metric $E({\bm{X}})$ is 10, so the radius $\beta$ is reduced from infinity to 10. In the forth and fifth steps the accumulated partial metrics $E({\bm{X}}_{k:N - 1})$ are larger than $\beta$, so the nodes and their child nodes are discarded. In the eighth step the new eligible decision $\bm{X}$ with smaller metric is obtained, the radius $\beta$ is reduced from 10 to 9. In the ninth and tenth steps, the nodes and their child nodes are also discarded because their accumulated partial metrics $E({\bm{X}}_{k:N - 1})$ are larger than $\beta$. Then, all the possible decisions have been searched and the search process stops.

\section{Combination of Sphere Decoder \\with Box Optimization} \label{section4}
As we known, the number of expanded nodes in the search process is a key factor related to complexity. Compared with MLD, although the conventional SD described in Section \ref{section3} reduces the complexity by discarding some child nodes when the inequality is not satisfied, the search process is still complex. The combination of SD and BO can further reduce the number of expanded nodes, so that the complexity of receiver can be further reduced.

\subsection{Finding the Optimal Decision by SD with BO}
The process to find the optimal decision by SD with BO consists of two steps, the first is to get the initial solution and set the initial radius $\beta$. A continuous unconstrained least-square problem is usually solved to get the initial solution,
\begin{equation}
{\bm{S}} = \mathop {\arg \min }\limits_{{\bm{X}}\in {\mathbb{C}^N}} {\left\| {{\bm{Z} - \bm{RX}}} \right\|^2} = \mathop {\arg \min }\limits_{{\bm{X}}\in {\mathbb{C}^N}} E(\bm{X}),
\label{eq13}
\end{equation}
where $\mathbb{C}$ means the set of complex numbers. Then, the Eq. (\ref{eq13}) is solved by
\begin{equation}
\bm{S} = \bm{R} \backslash \bm{Z}.
\label{eq14}
\end{equation}
In Eq. (\ref{eq14}), the backslash ``$\backslash$'' is a MATLAB operator, which is accomplished by the Moore-Penrose pseudo-inverse (defined by $(\cdot)^+$) as $\bm{R} \backslash \bm{Z}=\bm{R}^{+}\bm{Z}=(\bm{R}^T\bm{R})^{-1}\bm{R}^T\bm{Z}$ \cite{ZF}.

The solution $\bm{S}$ is said to be in the constellation when all the elements satisfy
\begin{equation}
\min (\Re(D)) \le {\Re(S_k)} \le \max (\Re(D)), 0\le k \le N-1,
\label{eq15}
\end{equation}
\begin{equation}
\min (\Im(D)) \le {\Im(S_k)} \le \max (\Im(D)), 0\le k \le N-1.
\label{eq16}
\end{equation}
If $\bm{S}$ is in the constellation, the initial solution ${\bm{S}}_{zf}$ can be the nearest constellation of $\bm{S}$, which is called zero-forcing estimator, and the initial radius is set to $\beta  = \left\| {{\bm{Z}} - {\bm{R}}{{\bm{S}}_{zf}}} \right\|^2$. When the noise and ICI seriously degrade the received signal, Eqs. (\ref{eq15}) and (\ref{eq16}) are not satisfied, so $\bm{S}$ is said to be out of the constellation. Then, the BO algorithm is used to obtain the initial solution ${\bm{S}}_{bo}$, and the initial radius is set to be $\beta  = \left\| {{\bm{Z}} - {\bm{R}}{{\bm{S}}_{bo}}} \right\|^2$. The details of BO will be described later.


\begin{algorithm}[!t]
\caption{Search process} 
\begin{algorithmic}[1]
\For{$X_k \in \mathcal{D}$ with S-E enumeration}
    \If{Eq. (\ref{eq12}) is satisfied} 
　　　　\State Calculate the solution ${{\bm{S}}_{0:k - 1}}$ as Eq. (\ref{eq19}).
　　　　\If {${{\bm{S}}_{0:k - 1}}$ is in the constellation}
            \State Expand the current node of subcarrier $k$.
        \Else
            \State Use BO to obtain ${{\bm{S}}_{0:k - 1}^{bo}}$.
            \State Calculate $\delta$ as Eq. (\ref{eq20}).
            \If {Eq. (\ref{eq17}) is satisfied}
                \State Expand the current node of subcarrier $k$.
            \Else
                \State Discard the current node of subcarrier $k$.
            \EndIf
        \EndIf
　　\Else
　　　　\State Discard the current node of subcarrier $k$.
        \State Exit for loop.
　　\EndIf
\EndFor
\end{algorithmic}
\label{alg1}
\end{algorithm}

The second step is to find the decision $\bm{X}$ with minimal metric. The S-E enumeration is also applied in the search process of SD with BO. After making decision for subcarrier $k$, the first term $E({\bm{X}}_{0:k - 1})$ in Eq. (\ref{eq11}) can be utilized to obtain a bound $\delta$, so the Eq. (\ref{eq12}) is expressed with a tighter bound as
\begin{equation}
\resizebox{.89\hsize}{!}{${E({\bm{X}}_{k:N - 1})}\! =\! {\left\| {{{\bm{Z}}_{k:N - 1}} \!-\! {{\bm{R}}_{k:N - 1,k:N - 1}}{{\bm{X}}_{k:N - 1}}} \right\|^2} \! \le \! \beta \! - \! \delta$}.
\label{eq17}
\end{equation}
To get the bound $\delta$, we should first solve the ${{\bm{S}}_{0:k - 1}}$ by
\begin{equation}
\resizebox{.89\hsize}{!}{$\begin{array}{l}
{{\bm{S}}_{0:k - 1}} \!=\! \mathop {\arg \min }\limits_{{{\bm{X}}_{0:k-1}} \in {\mathbb{C}^{k}}} \left\| {{{\bm{Z}}_{0:k - 1}} \!-\! {{\bm{R}}_{0:k - 1,0:k - 1}}{{\bm{X}}_{0:k - 1}}} \right.\\
{\kern 110pt}{ - {\bm{R}}_{0:k - 1,k:N - 1}}{\left. {{{\bm{X}}_{k:N - 1}}} \right\|^2}.
\end{array}$}
\label{eq18}
\end{equation}
The method to obtain the solution ${{\bm{S}}_{0:k - 1}}$ is just a deflated problem of getting the initial solution ${\bm{S}}$ in Eq. (\ref{eq13}). ${{\bm{S}}_{0:k - 1}}$ is calculated by
\begin{equation}
\resizebox{.89\hsize}{!}{${{\bm{S}}_{0:k - 1}} \!=\! {\bm{R}}_{0:k - 1,0:k - 1}\backslash({\bm{Z}}_{0:k - 1}\!-\!{\bm{R}}_{0:k - 1,k:N - 1}{\bm{X}}_{k:N - 1})$}.
\label{eq19}
\end{equation}
If ${{\bm{S}}_{0:k - 1}}$ is in the constellation, the bound $\delta$ is no longer taken into consideration and we directly expand the node of subcarrier $k$. If ${{\bm{S}}_{0:k - 1}}$ is out of the constellation, ${{\bm{S}}_{0:k - 1}}$ is utilized as the initial solution of BO, and BO is applied to obtain the solution ${{\bm{S}}_{0:k - 1}^{bo}}$, so the bound $\delta$ is expressed as
\begin{equation}
\resizebox{.89\hsize}{!}{$\delta \!=\! \left\| {{{\bm{Z}}_{0:k - 1}} \!-\! {{\bm{R}}_{0:k - 1,0:k - 1}}{{\bm{S}}_{0:k - 1}^{bo}} \!-\! } \right.{{\bm{R}}_{0:k - 1,k:N - 1}}{\left. {{{\bm{X}}_{k:N - 1}}} \right\|^2}$}.
\label{eq20}
\end{equation}
Then we can use the bound $\delta$ to decide whether the node of subcarrier $k$ should be expanded. The whole search process of SD with BO is summarized in Algorithm \ref{alg1}.

\subsection{Details of BO algorithm}
The BO algorithm only can be applied to real-valued signal. If the initial solution inputting into BO is a complex signal, the real part and imaginary part are handled separately and combined at last. The concepts of active set $\mathcal{B}$ and free set $\mathcal{F}$ are introduced into BO algorithm. For $\bm{X} \in \mathbb{R}^k$ ($\mathbb{R}$ means the set of real numbers), the index set $i$ of $\bm{X}$ can be divided into
\begin{equation}
\left\{ {0,1,\dots} \right\}=\mathcal{B}\cup\mathcal{F}.
\label{eq21}
\end{equation}
For $i \in \mathcal{B}$, $X_i$ is fixed to upper bound or lower bound of the constellation set (i.e., $\max (\Re(D))$, $\min (\Re(D))$, $\max (\Im(D))$ and $\min (\Im(D))$. For all the other $i \in \mathcal{F}$, $X_i$ is just the free variable, so $\mathcal{F}$ is called the free set. The process of BO algorithm is shown as follows \cite{BO,ActiveSetAlg},

\begin{enumerate}[1)]
\item Preprocessing: Start from the initial solution ${\bm{X}}^0$, divide ${\bm{X}}^0$ into active set ${\bm{X}}_\mathcal{B}^0$ and free set ${\bm{X}}_\mathcal{F}^0$.
\item FSM: Solve the least-square problem for the elements in the free set,
\begin{equation}
{\bm{X}}_\mathcal{F}'={\bm{R}}_\mathcal{F}\backslash({\bm{Z}}-{\bm{R}}_\mathcal{B}{\bm{X}}_\mathcal{B}^0),
\label{eq22}
\end{equation}
where ${\bm{R}}_\mathcal{B}={\bm{R}}(:,\mathcal{B})$ and ${\bm{R}}_\mathcal{F}={\bm{R}}(:,\mathcal{F})$.
\begin{enumerate}[a)]
\item If ${\bm{X}}'= {\bm{X}}_\mathcal{B}^0\cup{\bm{X}}_\mathcal{F}'$ is in the constellation, calculate the Lagrange multiplier ${\bm{L}}$ of ${\bm{X}}'$ as
\begin{equation}
{\bm{L}}={\bm{R}}^T({\bm{Z}}-{\bm{R}}{\bm{X}}').
\label{eq23}
\end{equation}
\begin{itemize}
\item If ${\bm{L}}$ is optimal, the optimal solution ${\bm{X}}'$ is obtained and the BO algorithm is finished;
\item If ${\bm{L}}$ is not optimal, get the element whose multiplier deviates most from optimality, move it from $\mathcal{B}$ to $\mathcal{F}$. Set ${\bm{X}}^0={\bm{X}}'$ and go back to FSM.
\end{itemize}

\item If ${\bm{X}}'= {\bm{X}}_\mathcal{B}^0\cup{\bm{X}}_\mathcal{F}'$ is out of the constellation, the descent direction and step size should be calculated to let the new solution have the possibility to be in the constellation. The descent direction is calculated as ${\bm{X}}'-{\bm{X}}^0$, the step size is calculated by
\begin{equation}
{t_i} = \left\{ \begin{array}{l}
(X_i^0 - l)/(X_i^0 - X_i'),X_i' < l\\
(u - X_i^0)/(X_i' - X_i^0),X_i' > u,
\end{array} \right.
\label{eq24}
\end{equation}
where $i \in \mathcal{F}$, $l=\min(\Re(D))$ or $l=\min(\Im(D))$, $u=\max(\Re(D))$ or $u=\max(\Im(D))$. Then, the step size is $t = \mathop {\min }\limits_{i \in F} {t_i}$, and the new initial solution is
\begin{equation}
{\bm{X}}^0={\bm{X}}^0+t({\bm{X}}'-{\bm{X}}^0).
\label{eq25}
\end{equation}
    After getting new initial solution ${\bm{X}}^0$, at least one element is moved from $\mathcal{F}$ to $\mathcal{B}$. Then, go back to FSM.
\end{enumerate}
\end{enumerate}

In BO algorithm, the element ${X_i}$ exchanges between $\mathcal{F}$ and $\mathcal{B}$ until the Lagrange multiplier ${\bm {L}}$ is optimal. The Lagrange multiplier ${\bm {L}}$ is said to be optimal that for all $i\in \mathcal{B}$, $L_i\le 0$ when ${X_i}$ is fixed to upper bound of the constellation set, and $L_i\ge 0$ when ${X_i}$ is fixed to lower bound of the constellation set \cite{Lagrange}. The solution $\bm{X}'$ is bounded in a zone like a box as Eqs. (\ref{eq15}) and (\ref{eq16}), hence this algorithm is called BO algorithm.

\section{Simulation Results} \label{section5}
In the simulation, the performance of FTN-NOFDM system is evaluated in AWGN channel. The subcarrier number is set to 16. BER against $E_b/N_0$ is commonly used to compare the performance of different schemes. The noise power $P_n$ of the received signal is calculated by the signal-to-noise ratio (SNR). The relationship between SNR (i.e., $P_s/P_n$) and $E_b/N_0$ can be expressed as
\begin{equation}
\frac{{{P_s}}}{{{P_n}}} = \frac{{{E_s}}}{{{N_0}{B_n}{T_s}}} = \frac{{{E_s}{R_s}}}{{{N_0}{B_n}}} = \frac{{{E_s}}}{{{N_0}\alpha}} = \frac{{{E_b}{\log_2M }}}{{{N_0}\alpha}},
\label{eq26}
\end{equation}
where $P_s$ is the average signal power and $E_s$ is the energy per symbol of the transmitted FTN-NOFDM signal. $P_n$ is the noise power and $B_n$ is the noise bandwidth. The symbol rate $R_s$ is equal to $1/T_s$. In FTN-NOFDM system, the transmission rate is faster than the Nyquist rate, so that $R_s/B_n=1/\alpha$. The ${\log_2M }$ indicates the number of bits per symbol including for $M$-QAM modulation.

When the conventional SD and SD with BO are applied to reduce ICI, it should be noted that the correlation matrix $\bm{C}$ represented the ICI is a real-valued matrix. In general, the decision of symbol on the $k$-th subcarrier (i.e., $X_k$) employs the complex constellation set directly. Owing to the real-valued property of ICI matrix $\bm{C}$, the decision of $X_k$ can also be separated to the decision of real part and imaginary part of $X_k$ independently. The computational time of complex search directly for the conventional SD in FTN-NOFDM is too long to make meaningful simulation, so we only make comparison of complex search directly and real/imaginary (Re/Im) search separately for SD with BO in FTN-NOFDM. The simulation results show that these two search methods for SD with BO have entirely the same performance to reduce ICI.

\begin{figure}[!t]
\centering
\includegraphics[width=7.1cm]{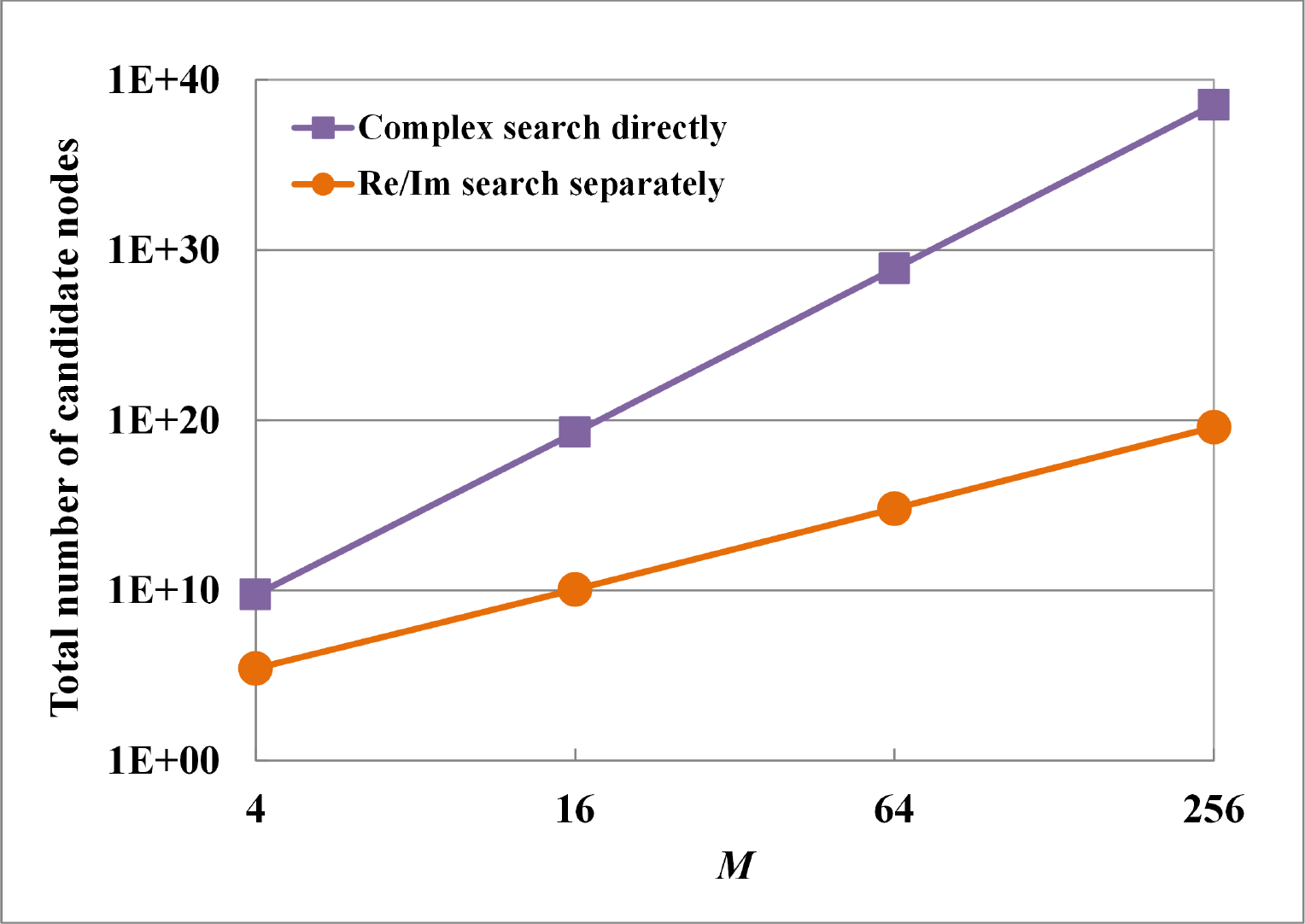}
\caption{Total number of candidate nodes for complex search directly and Re/Im search separately in FTN-NOFDM with $M$-QAM modulation and 16 subcarriers.}
\label{fig5}
\end{figure}

The number of expanded nodes after decoding all the subcarriers is commonly used to represent the complexity of these two search methods. The expanded nodes are chosen from the candidate nodes. The total number of candidate nodes for complex search directly in FTN-NOFDM with $M$-QAM modulation and $N$ subcarriers is
\begin{equation}
M+M^2+\dots+M^N = \frac{{{M(1-M^N)}}}{{{1-M}}}.
\label{eq27}
\end{equation}
The total number of candidate nodes for Re/Im search separately in FTN-NOFDM with $M$-QAM modulation and $N$ subcarriers is
\begin{equation}
2\times(\sqrt M+\sqrt M^2+\dots+\sqrt M^N) = \frac{{{2\sqrt M(1-\sqrt M^N)}}}{{{1-\sqrt M}}}.
\label{eq28}
\end{equation}
Fig. \ref{fig5} illustrates the difference between complex search directly and Re/Im search separately more clearly. The total number of candidate nodes is calculated by Eqs. (\ref{eq27}) and (\ref{eq28}) with $M=4,16,64,256$ and $N=16$. It can be seen that the difference between the total number of candidate nodes for complex search directly and Re/Im search separately is increased when the high-order modulation format is used.

\begin{figure}[!t]
\centering
\includegraphics[width=7cm]{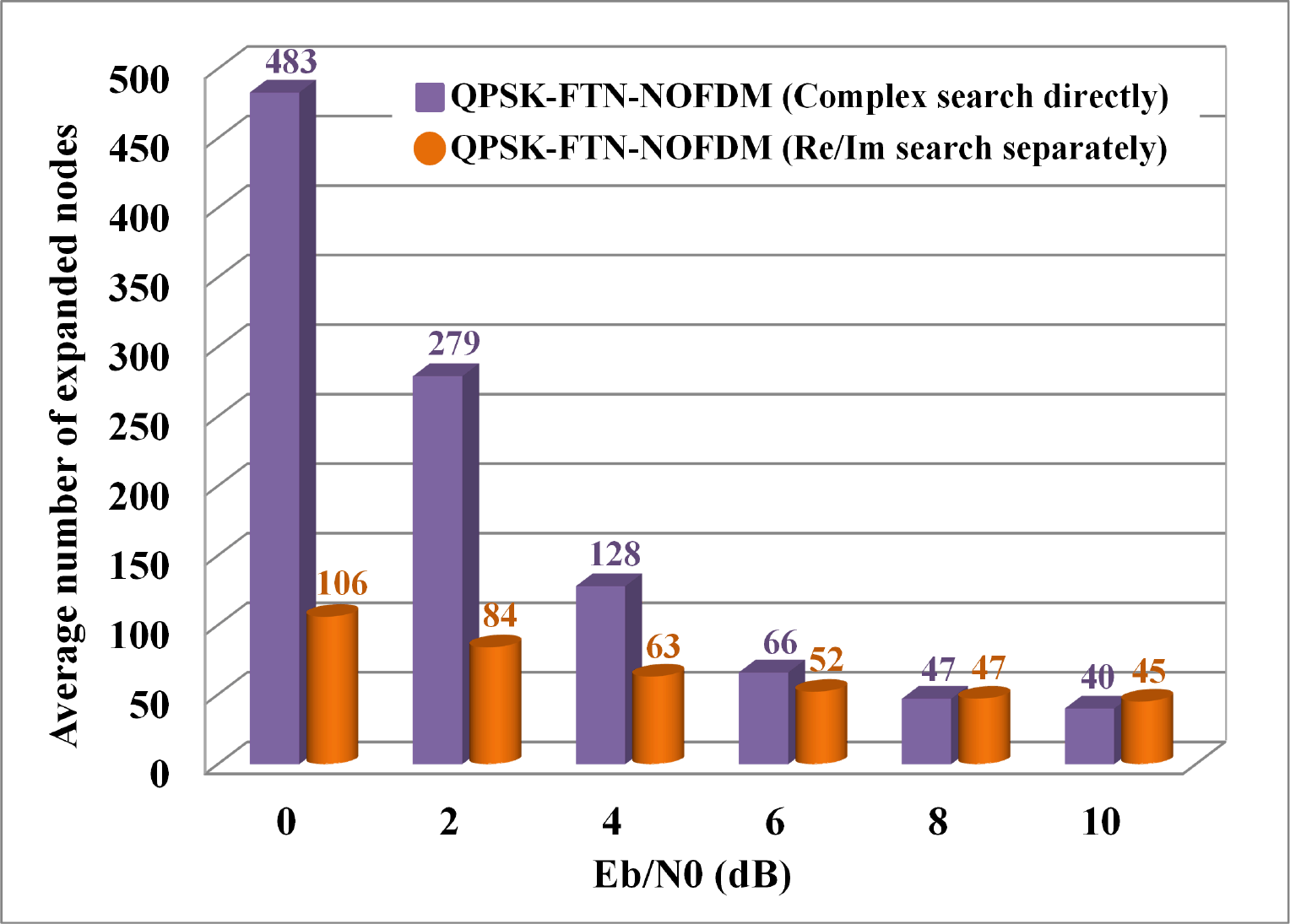}
\caption{Average number of expanded nodes for complex search directly and Re/Im search separately in SD with BO. QPSK modulation is used for FTN-NOFDM system with $\alpha=0.802$.}
\label{fig6}
\end{figure}

\begin{figure}[!t]
\centering
\includegraphics[width=7cm]{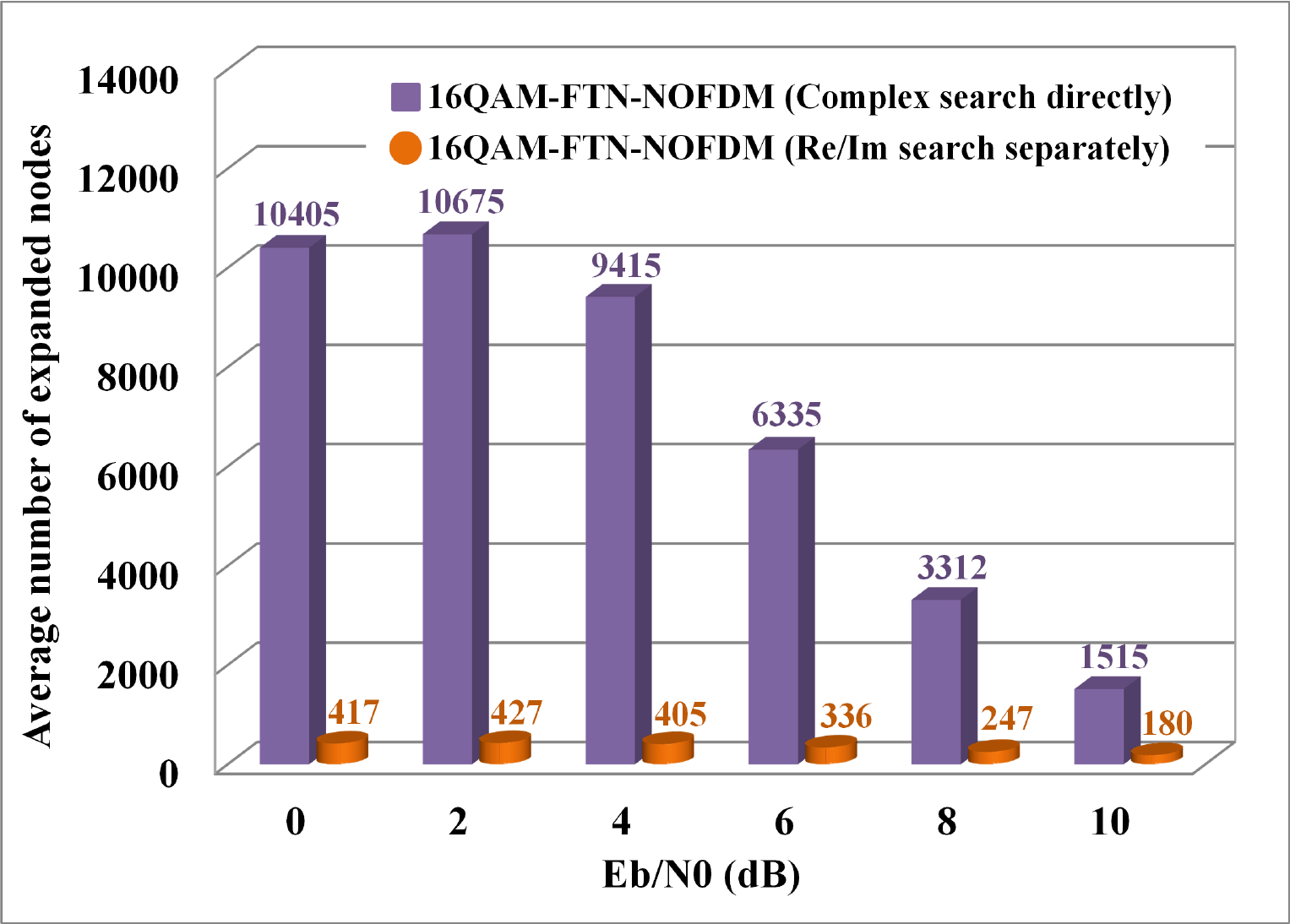}
\caption{Average number of expanded nodes for complex search directly and Re/Im search separately in SD with BO. 16QAM modulation is used for FTN-NOFDM system with $\alpha=0.802$.}
\label{fig7}
\end{figure}

Figures \ref{fig6} and \ref{fig7} demonstrate the average number of expanded nodes of complex search directly and Re/Im search separately for SD with BO. The bandwidth compression factor $\alpha$ is set to 0.802 as an example. For QPSK modulation, the average number of expanded nodes for complex search directly is larger than that of Re/Im search separately when the $E_b/N_0$ is smaller than 8. When the $E_b/N_0$ is equal or larger than 8, it becomes almost the same for two search methods. The reason is that when the noise is small, the radius $\beta$ is reduced quickly and the path with minimal metric can be found without expanding so many nodes. For 16QAM modulation, the average number of expanded nodes for Re/Im search directly is greatly reduced, because the difference between the total number of candidate nodes for complex search directly and Re/Im search separately is greatly larger than that for QPSK modulation. The Re/Im search separately can reduce the complexity especially for high-order modulation format, and it has the same performance to reduce ICI as complex search directly. Therefore, we use Re/Im search separately for both the conventional SD and SD with BO in the following simulations. When solving Eq. (\ref{eq13}) and Eq. (\ref{eq18}), the solution is chosen from the set of real numbers instead of the set of complex numbers.

\begin{figure}[!t]
\centering
\includegraphics[width=6cm]{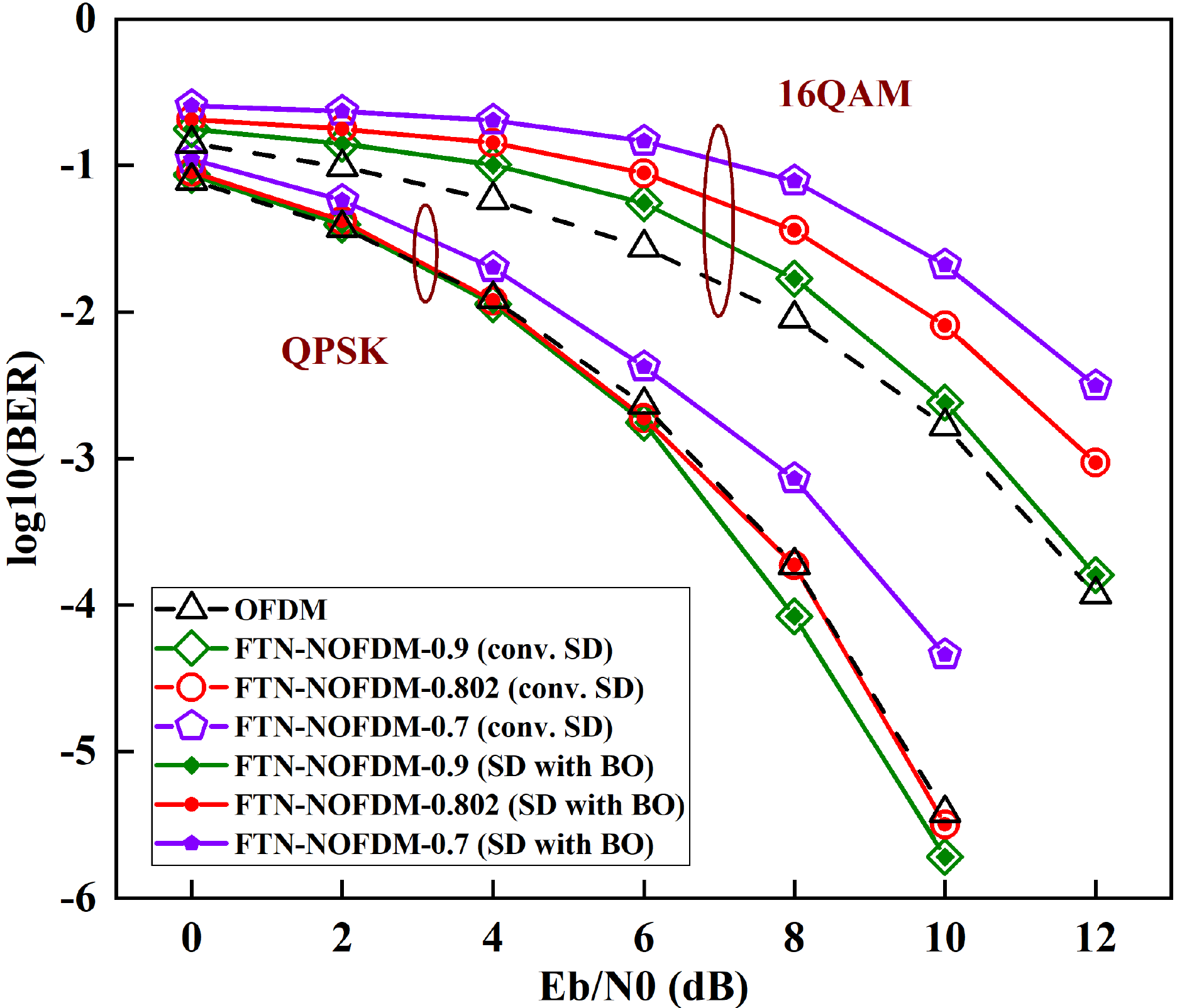}
\caption{BER performance of FTN-NOFDM system and OFDM system, the conventional SD and SD with BO are used in FTN-NOFDM system.}
\label{fig8}
\end{figure}

Figure \ref{fig8} illustrates that for QPSK-modulated and 16QAM-modulated FTN-NOFDM, the conventional SD and SD with BO have the same performance to reduce ICI. For QPSK modulation, FTN-NOFDM with $\alpha$ equal to 0.802 has almost the same performance as OFDM, which agrees well with the Mazo limit. The BER performance of FTN-NOFDM with $\alpha$ equal to 0.9 is even a little bit better than that of OFDM, the reason can be found from Eq. (\ref{eq26}) that the noise is distributed on all the bandwidth, but the effective noise bandwidth that affects the signal should multiply $\alpha$ owing to the compression of signal bandwidth. The results also verify the capacity limit of FTN-NOFDM signal, that when $\alpha$ is between 1 and 0.802, the ICI can be effectively eliminated by an optimal detector, then the capacity limit of FTN-NOFDM signal can be higher than that of Nyquist signal \cite{FrCT-FTN}. Moreover, the BER performance of 16QAM-modulated FTN-NOFDM degrades once $\alpha$ is less than 1. Therefore, more effective algorithm to reduce ICI in FTN-NOFDM with high-order constellation can be investigated further.

\begin{figure}[!t]
\centering
\includegraphics[width=6cm]{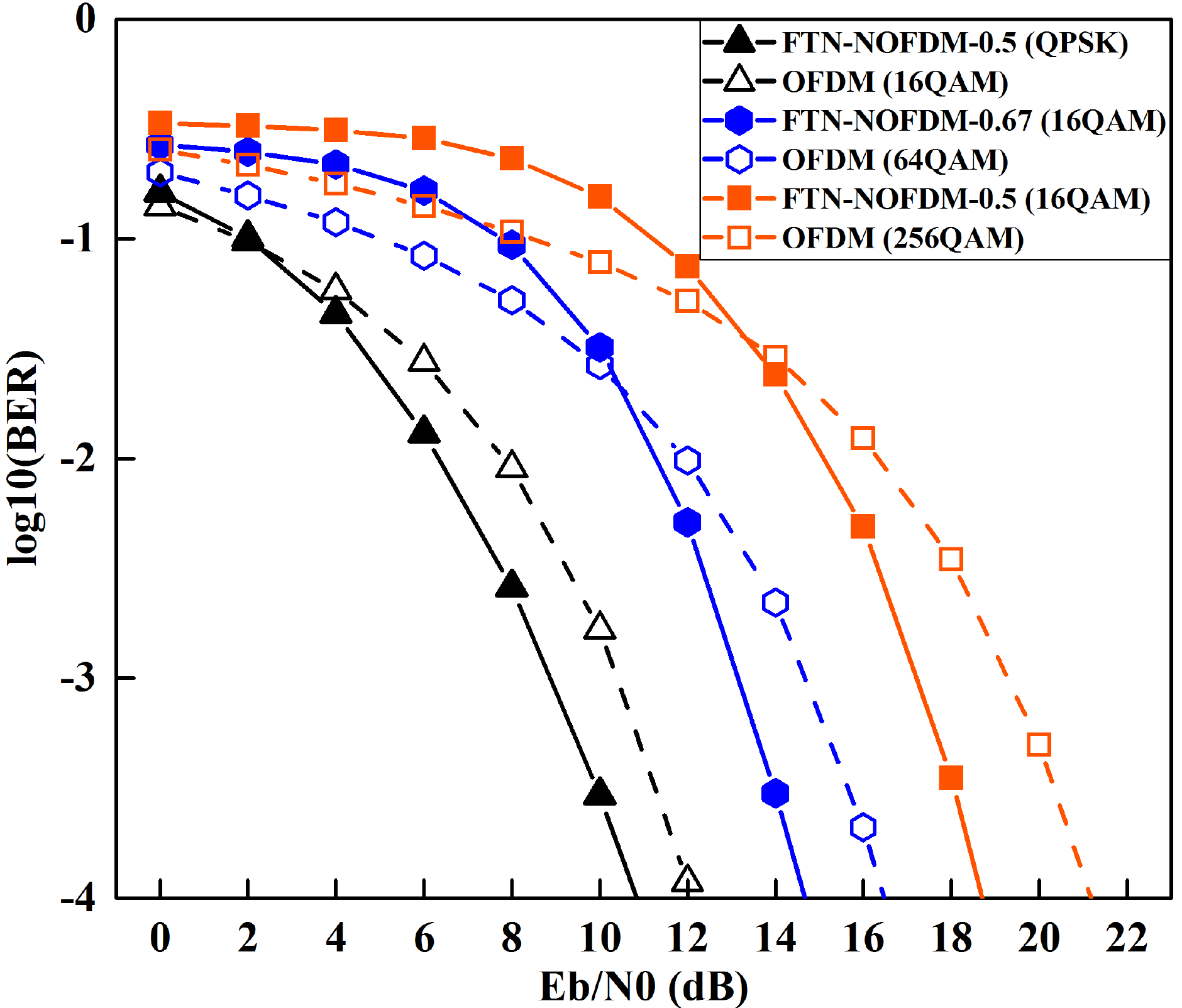}
\caption{BER performance comparison between FTN-NOFDM system and OFDM system with the same spectral efficiency, the spectral efficiency is 4 bit/s/Hz, 6 bit/s/Hz and 8 bit/s/Hz, respectively.}
\label{fig9}
\end{figure}

Under the same spectral efficiency, the BER performance comparison between FTN-NOFDM and OFDM is illustrated in Fig. \ref{fig9}. The BER performance curves of FTN-NOFDM and its counterpart OFDM with the same spectral efficiency have a cross, because the detector can not effectively reduce ICI when the noise is large. Nevertheless, we always pay more attention to the $E_b/N_0$ for a suitable BER such as $10^{-3}$. At the BER of $10^{-3}$, the QPSK-modulated FTN-NOFDM with $\alpha$ equal to 0.5 (the spectral efficiency is 4 bit/s/Hz) performs better than 16QAM-modulated OFDM with about 1.5 dB $E_b/N_0$ gain. The 16QAM-modulated FTN-NOFDM with $\alpha$ equal to 0.67 and 0.5 (the spectral efficiency is 6 bit/s/Hz and 8 bit/s/Hz, respectively) has about 1.5 dB and 2 dB $E_b/N_0$ gain at the BER of $10^{-3}$ compared with 64QAM-modulated and 256QAM-modulated OFDM, respectively. Therefore, the QPSK-modulated or 16QAM-modulated FTN-NOFDM has better BER performance than OFDM with high-order constellation under the same spectral efficiency.

\begin{figure}[!t]
\centering
\includegraphics[width=7cm]{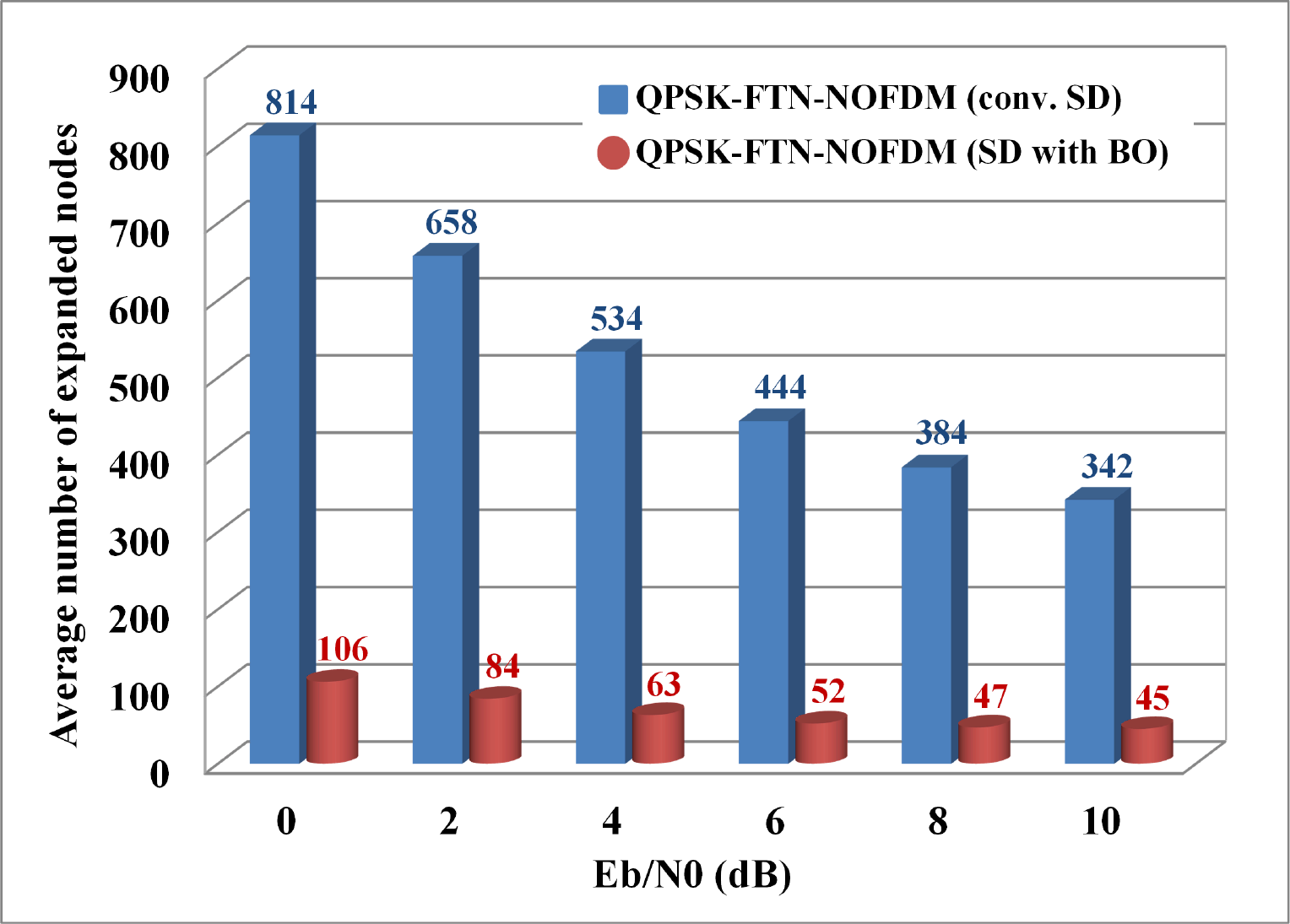}
\caption{Average number of expanded nodes of the conventional SD and SD with BO for FTN-NOFDM system, the QPSK modulation is used and $\alpha$ is equal to 0.802.}
\label{fig10}
\end{figure}

\begin{figure}[!t]
\centering
\includegraphics[width=7cm]{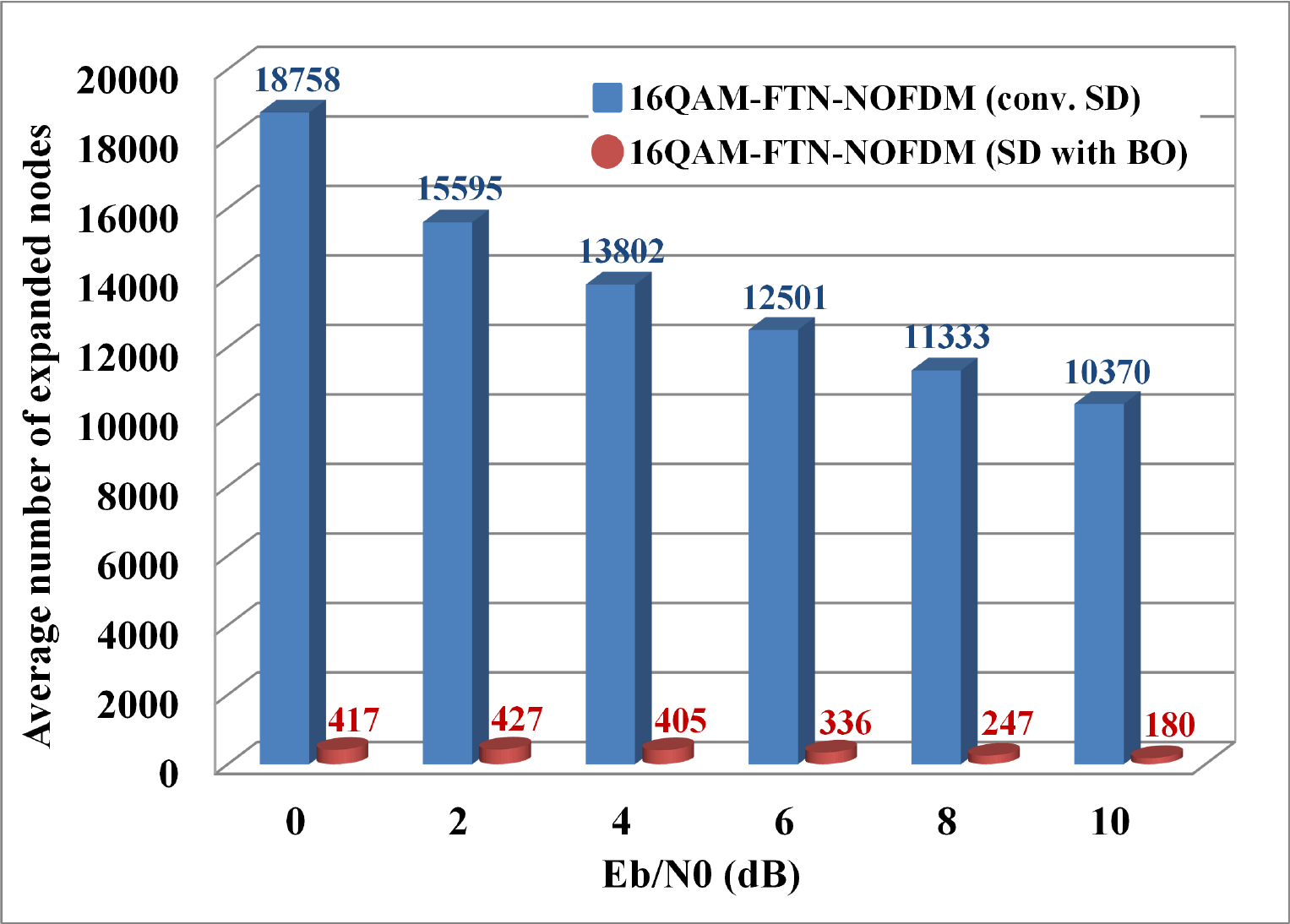}
\caption{Average number of expanded nodes of the conventional SD and SD with BO for FTN-NOFDM system, the 16QAM modulation is used and $\alpha$ is equal to 0.802.}
\label{fig11}
\end{figure}

Figures \ref{fig10} and \ref{fig11} show the average number of expanded nodes of the conventional SD and SD with BO for FTN-NOFDM system with QPSK modulation and 16QAM modulation, respectively. The bandwidth compression factor $\alpha$ is set to 0.802 as an example, and the Re/Im search separately is used for both the conventional SD and SD with BO. The average number of expanded nodes has the tendency to decrease with the increasing of $E_b/N_0$. Compared with the conventional SD, the average number of expanded nodes for SD with BO is significantly decreased, and this decrease is more obvious for 16QAM modulation than for QPSK modulation. Therefore, SD with BO has lower complexity than the conventional SD, and SD with BO can significantly reduce the complexity for high-order modulation format.

\section{Conclusion} \label{section6}
In this paper, we apply SD with BO to reduce ICI in FTN-NOFDM system. SD with BO has the same performance to reduce ICI as the conventional SD. Its average number of expanded nodes in search process is significantly decreased especially for high-order modulation format, which can reduce the complexity of the receiver. The QPSK-modulated FTN-NOFDM agrees well with the Mazo limit, of which the transmission rate can achieve 24.7\% faster than its Nyquist rate. Under the same spectral efficiency, the BER performance of 16QAM-modulated FTN-NOFDM is better than that of OFDM with high-order constellation. In the future work, the more effective algorithm will be investigated for FTN-NOFDM with high-order constellation. In conclusion, FTN-NOFDM will be a promising modulation scheme for the future bandwidth-limited wireless communications.

\ifCLASSOPTIONcaptionsoff
  \newpage
\fi



%

\end{document}